\documentclass[12pt,epsfig,citesort,svgnames,dvipsnames]{article}
\usepackage{multicol}
\usepackage[all]{xy}
\usepackage{graphicx}
\usepackage{caption,subcaption}
\usepackage{todonotes}
\usepackage{amsmath}
\usepackage{amssymb}
\usepackage{amsfonts}
\usepackage{amscd}
\usepackage{latexsym}
\usepackage[official]{eurosym}
\usepackage[english]{babel}
\usepackage{csquotes}
\usepackage{authblk}
\usepackage{color}
\usepackage{url}
\usepackage{enumitem}
\usepackage{xfrac}

\usepackage{natbib}
\usepackage{xcolor}
\usepackage{hyperref}\usepackage[nottoc]{tocbibind}
\hypersetup{colorlinks=true,citecolor=WildStrawberry,anchorcolor=Black,filecolor=cyan,menucolor=magenta,runcolor=cyan,linkcolor=RoyalBlue,linktoc=page, urlcolor=TealBlue}

\usepackage[T1]{fontenc}
\usepackage{newpxtext,newpxmath}

\usepackage[a4paper,margin=1in]{geometry}
\usepackage[protrusion=true,expansion=true]{microtype}

\include{head}

\title{\textbf{A de Broglie-Bohm Model of Pure Shape Dynamics: \\$N$-body system}}
\author[a,b]{\normalsize Pooya Farokhi\thanks{\href{mailto:pooyafarokhi.p@gmail.com}{pooyafarokhi.p@gmail.com}}}
\author[c]{\normalsize Tim Koslowski\thanks{\href{mailto:t.a.koslowski@gmail.com}{t.a.koslowski@gmail.com}}}
\author[d]{\normalsize Pedro Naranjo\thanks{\href{mailto:pnpfisica@gmail.com}{pnpfisica@gmail.com}}}
\author[e]{\normalsize Antonio Vassallo
\thanks{\href{mailto:antonio.vassallo1977@gmail.com}{antonio.vassallo1977@gmail.com}}}
\affil[a]{\normalsize \emph{Department of Physics, Sharif University of Technology, P.O. Box 11365-9161,} \break \emph{\normalsize Tehran, Iran}}
\affil[b]{\normalsize \emph{University of Cologne, Department of Physics, Albertus-Magnus-Platz,} \break \emph{\normalsize 50923 Cologne, Germany}}
\affil[c]{\normalsize  \emph{Technical University of Applied Sciences W\"urzburg-Schweinfurt, Faculty of Applied Natural Sciences and Humanities, M\"unzstr. 12, 97070 W\"urzburg, Germany}}
\affil[d]{\normalsize \emph{Plaza Mayor, 4/1B, 09003 Burgos, Spain}}
\affil[e]{\normalsize \emph{Warsaw University of Technology, Faculty of Administration and Social Sciences,\break Plac Politechniki 1, 00-661 Warsaw, Poland}}

\date{}

\begin{document}
\maketitle

\begin{abstract}
  \noindent We provide the construction of a de Broglie-Bohm model of the $N$-body system within the framework of Pure Shape Dynamics. The equation of state of the curve in shape space is worked out, with the instantaneous shape being guided by a wave function. In order to get a better understanding of the dynamical system, we also give some numerical analysis of the 3-body case. Remarkably enough, our simulations typically show the attractor-driven behaviour of complexity, well known in the classical case, thereby providing further evidence for the claim that the arrow of complexity is the ultimate cause of the experienced arrow of time.
\end{abstract}
\newpage
\tableofcontents

\section{Introduction}

The term ``Quantum gravity'' has evolved into a vast umbrella term for a host of different, many mutually incompatible, approaches to a theory that successfully reconciles the principles of classical gravity and quantum theory. Despite almost a century of intense efforts, no such theory exists as of yet. One might further claim that there is no consensus regarding even what this putative theory of ``quantum gravity'' should mean. We believe a serious rethinking is called for already at the level of \emph{classical} gravity. 

As is well-known by now, Shape Dynamics (SD) offers a description of classical gravity in terms of an evolving conformal structure, unlike the standard spacetime description of General Relativity (GR) (see \citealp{390}, for a nice introduction to the subject, with an emphasis on conceptual matters, \citealp{514}, for a pedagogical, yet comprehensive, account, and \citealp{528}, for the Hamiltonian version, which renders the mathematical relation between GR and SD transparent). More precisely, SD is concerned with describing the geometry of the curve traced out by the relevant physical system in relational configuration space, known as \emph{shape space}, whereby this curve describes the entire history of objective relations between subsystems within said physical system. At its finest, displaying its Machian core, SD is a theory of the whole universe: the curve in shape space contains all objective statements about the universe and its history.

As a first step towards a theory of quantum gravity, in this paper we shall concern ourselves with a much more humble task: to describe a minimal model of the quantum $N$-body system within the general framework of Pure Shape Dynamics (PSD). It is thus the natural continuation of the classical setting provided in \citet{737}. This is achieved within the general framework of de Broglie-Bohm theory, suitably adapted to comply with our relational tenets (\citealp{338,187a,187b}, give the original papers, \citealp{360}, offers a comprehensive account and \citealp{235}, describes the somewhat different theory referred to as ``Bohmian mechanics''). Accordingly, the physical objects of our model are $N$ point-like particles, with the ratios of their separations defining the associated shape space. As mentioned above, the dynamics of the shape of this $N$-body universe is given by the curve traced out in said space. We should like to emphasise the ``architectonic features'' shared by the relational $N$-body system and dynamical geometry, whereby the structural similarity between the two theories enables us to address \emph{the} key issue at the heart of quantum gravity already in the simple $N$-body system: the notorious problem of time is given a novel perspective if we consider that it is the arrow of \emph{complexity} that defines the arrow of time \citep{712,706}. 

As already mentioned, in order to provide a purely relational, background-independent account of quantum phenomena, we shall consider the PSD version of the de Broglie-Bohm $N$-body model. Our choice of this formulation of quantum theory is twofold: it i) accords to spatial structures a privileged status, much like SD does (the former in terms of instantaneous positions, and the latter through instantaneous conformal structure), and ii) provides a simple solution to the infamous \emph{measurement problem} plaguing the conceptual understanding of quantum theory (see \citealp{197}, for a general assessment). 

This ``outcome'' problem, as we would like to refer to it, manifests itself in unitary quantum mechanics (QM) when one attempts to describe an isolated quantum system without an external classical measurement apparatus, as is the case for the universe as a whole. It is important to stress that the outcome problem has nothing to do with the statistical nature of the outcomes of measurements. Instead, it is the fundamental problem that unitary QM cannot produce statistical outcomes.

To look at this issue from another perspective, we should like to emphasise that this problem at the heart of unitary QM should be quite appropriately called the ``reality'' problem, for it is more about positing a set of objectively stable and determinate physical entities---so-called 
\emph{beables}, after John Bell's coining \citep{195}---and aiming for an objective physical description of quantum phenomena in terms of these beables. This enterprise is simply dismissed as erroneous in the standard framework of QM. The de Broglie-Bohm theory is a member of the family of approaches that instead allows for such an objective description of quantum theory and is singled out within this family as the member that is structurally closest to the PSD framework.%Among the various formulations of quantum theory that do consider an objective description is the de Broglie-Bohm theory.

The de Broglie-Bohm theory describes the evolution of the universe as a curve in its configuration space, with this curve being ``guided'' by a wave equation. In order to be consistent with the relational nature of SD, we will consider the PSD version of de Broglie-Bohm: both the curve traced out by the universe and the guiding wave function are defined in shape space, leading to the equation of state of the unparametrized curve coupled with the wave function degrees of freedom, thus extending the analysis of the classical setting to the quantum realm.

The structure of this paper is as follows. \S\ref{sec:OutcomeProblem} deals with the outcome problem in unitary QM, with \S\ref{subsec:reviewOutcome} providing a critical review, with an emphasis on the well-known ``basis problem'', showing that, contrary to some claims in the literature, although ``decoherence'' is a dominant phenomenological effect, it does not lead to ``einselection'', which would, in turn, solve the outcome problem. \S\ref{subsec:basisproblem} shows how the de Broglie-Bohm theory does yield definite outcomes. Next, \S\ref{sec:dBBPSD} analyses the construction of the PSD version of the de Broglie-Bohm $N$-body system for an arbitrary potential, stressing how our model differs from its standard, non-relational counterpart (\S\ref{subsec:quantumNbody}). In \S\ref{subsec:scale} we shall discuss the important case of the Newtonian potential, which will allow us to clearly spell out the role of scale in our model as well as describe the classical limit in \S\ref{subsec:classicalNbody}, matching the findings of \citet{737}. In \S\ref{subsec:bornstat} we shall tackle the important question of the recovery of Born statistics for subsystems, already pointed out at the end of \S\ref{subsec:basisproblem} and brought up more fully in \S\ref{subsec:scale}. Once the conceptual nuances and subtleties have been addressed, in order to provide a more robust analysis of our model, and given the well-known formidable analytical obstacles of the $N$-body system, we shall present in \S\ref{subsec:numerical} a numerical analysis of the equation of state of our de Broglie-Bohm model for the simple 3-particle case. Remarkably enough, our preliminary findings exhibit the crucial attractor-driven behaviour of complexity, thus providing further evidence for the fundamental nature of the arrow of complexity in our experienced arrow of time. Finally, \S\ref{sec:conclusions} will review the main results and address some open questions.

\section{Outcome Problem and de Broglie-Bohm Theory}\label{sec:OutcomeProblem}
 
\subsection{Review of the problem}\label{subsec:reviewOutcome}

The outcome problem of unitary QM appears when we want to describe the universe as a whole, which forces us to describe measurements as interactions between subsystems of the universe. For definiteness, let us consider a universe that consists of a macroscopic number $N$ of subsystems, which we will label by an index $i$. For simplicity, let us assume that each subsystem possesses only a finite number $d_i$ of distinct effectively accessible\footnote{This effective accessibility for subsystems is often generated through spatial confinement and the finiteness of locally available energy.} states, so the Hilbert space $\mathcal H_i$ has dimension $d_i$. After choosing a basis, we have a general observable $\hat A$ of subsystem $i$
\begin{equation}
  \hat A_i = \sum_{n=0}^{d_i-1} n\,|A_n\rangle\langle A_n|,
\end{equation}   
where the eigenstates $|A_n\rangle$ are orthonormal, $\langle A_n|A_m\rangle=\delta_{n,m}$. The conjugate shift operator $\hat B_i=\sum_{n=0}^{d_i-1} n\,|B_n\rangle\langle B_n|$ is diagonal in the dual basis
\begin{equation}
  |B_n\rangle = \frac{1}{\sqrt{d_i}}\sum_{m=0}^{d_i-1}e^{2 \pi i\frac{n\,m}{d_i}}|A_m\rangle.
\end{equation}
The name ``shift operator'' is due to the fact that for an integer $m$
\begin{equation}
  e^{-i\,m\,\hat B}|A_n\rangle = |A_{(n+m)\,\textrm{mod}\,d_i}\rangle.
\end{equation}
The state of the universe is given by the wave function
\begin{equation}
 |\psi\rangle = \sum_{n_1,\ldots,n_N=0}^{d_1-1,\ldots,d_N-1}\,a_{n_1\ldots n_N}\,|A_{n_1}\rangle\otimes\cdots\otimes |A_{n_N}\rangle,
\end{equation}
where a few tensor factors may describe global relations, while the vast majority describe individual subsystems. Hence, we can derive the expectation value of any observable $\hat{O}_i$ as
\begin{equation}
  \langle\psi|\hat{\mathbb I}_1\otimes\cdots\otimes\hat{\mathbb I}_{i-1}\otimes\hat{O}_i\otimes\hat{\mathbb I}_{i+1}\otimes\cdots\otimes \hat{\mathbb{I}}_N|\psi\rangle=\textrm{Tr}_{\mathcal{H}_i}\left(\hat{\rho}_i\,\hat{O}_i\right),
\end{equation}
where $\hat \rho_i$ denotes the reduced density matrix of subsystem $i$.

Using this notation, we can model an instantaneous measurement of observable $\hat A_i$ that is registered as a shift in the $A$-basis of subsystem $j$ by an interaction with Hamiltonian
\begin{equation}\label{equ:MeassurementInteraction}
 H_{int}=g\,\hat{\mathbb I}_1\otimes\cdots\otimes\hat{\mathbb I}_{i-1}\otimes\hat{A}_i\otimes\hat{\mathbb I}_{i+1}\otimes\cdots\otimes\hat{\mathbb I}_{j-1}\otimes\hat{B}_j\otimes\hat{\mathbb I}_{j+1}\otimes\cdots\otimes \hat{\mathbb{I}}_N,
\end{equation}
where we assume a very short interaction time, $\delta t\to 0$, while fixing the product $\frac{g\,\delta t}{\hbar}=1$, because this interaction evolves product states as
\begin{equation}
  |\chi_I\rangle\otimes|A_n\rangle\otimes|\chi_{II}\rangle\otimes|A_m\rangle\otimes|\psi_{III}\rangle \to |\chi_I\rangle\otimes|A_n\rangle\otimes|\chi_{II}\rangle\otimes|A_{(m+n)\,\textrm{mod}\,d_j}\rangle\otimes|\psi_{III}\rangle\,,
\end{equation}
where the $|\chi\rangle$'s denote the complementary part of subsystems, so the state $|A_n\rangle$ of subsystem $i$ is recorded as a shift of size $n$ in subsystem $j$. However, expressing the same interaction in the $B$-basis leads to a reverse measurement,
\begin{equation}
  |\chi_I\rangle\otimes|B_n\rangle\otimes|\chi_{II}\rangle\otimes|B_m\rangle\otimes|\psi_{III}\rangle \to |\chi_I\rangle\otimes|B_{(n-m)\,\textrm{mod}\,d_i}\rangle\otimes|\chi_{II}\rangle\otimes|B_m\rangle\otimes|\psi_{III}\rangle,
\end{equation}
in which the state $|B_m\rangle$ of subsystem $j$ is recorded as a shift in the state of subsystem $i$. Thus, the distinction between subsystem and apparatus depends on the basis in which we express the subsystem. 

This observation is an aspect of the well-known \emph{basis selection problem}, which has many additional facets. This problem necessarily arises when one tries to obtain definite outcomes from unitary evolution alone. The question is: ``In which basis are the definite outcomes selected?'' The problem is that the probabilities calculated using the Born rule depend on the selected basis, which readily follows from the fact that the expansion in a basis is linear, whereas the Born rule is a modulus squared. So whether the Born rule is applied in the $A$-basis or in the $B$-basis leads in general to distinct probability distributions.

The claim of the so-called ``einselection programme'' is that \emph{decoherence}, i.e., the weak interaction between the apparatus (in the previous example the subsystem $j$) and the environment, selects a preferred ``pointer'' basis (for good and general accounts of decoherence, see \citealp{431,685}; for a recent and accessible introduction to einselection and related ideas, see \citealp{740}). To simplify the presentation, let us change the notation and consider the entire environment of subsystem $j$ as a combined system on a $D^{N-1}:=\frac{\prod_{i=1}^N d_i}{d_j}$-dimensional Hilbert space $\mathcal H_{\text{env}}$, so the state $|\psi\rangle$ of the universe can be written as
\begin{equation}
 |\psi\rangle = \sum_{i=1}^{d_j}\sum_{I=1}^{D^{N-1}} a_{i,I}\,|A_i\rangle\otimes |E_I\rangle.
\end{equation}
Let us now consider a generic nonvanishing interaction between subsystem $j$ and its environment, which means that the universe evolves as
\begin{equation}
 |\psi(t)\rangle = \sum_{i=1}^{d_j}\sum_{I=1}^{D^{N-1}} a_{i,I}(t)\,|A_i\rangle\otimes |E_I\rangle.
\end{equation}
We are now in the position to discuss the typical ``size'' of the matrix elements of the reduced density matrix. 
The normalization of $|\psi\rangle$ means that $\sum_{i=1}^{d_j}\sum_{I=1}^{D^{N-1}} |a_{i,I}(t)|^2=1$, so the states can be identified with points on the $\Delta=(2\,\prod_{i=1}^N d_i)-1$-dimensional sphere, which can be parametrized using angles $\phi_1,\ldots,\phi_\Delta$ on the domain $[0,\pi]^{\Delta-1} \times [0,2\pi]$. In order to remain neutral about the basis, our notion of typicality will make use of the unique normalized $U(\frac{\Delta+1}{2})$-invariant Fubini-Study measure\footnote{Note that the action of $U(\frac{\Delta +1}{2})$ on states is equivalent to the action of $\text{SO}(\Delta + 1)$ on the $\Delta$-sphere, which leaves the Euclidean metric and its corresponding volume form invariant. The measure~\eqref{Uinvmeasure} comes from the induced background Euclidean metric on the sphere.}
\begin{equation}\label{Uinvmeasure} 
d\mu=\frac{\Gamma\left(\frac{\Delta+1}{2}\right)}{2\pi^{\frac{\Delta+1}{2}}}\sin^{\Delta-1}\phi_1\,d\phi_1\,\sin^{\Delta-2}\phi_2\,d\phi_2\cdots \sin\phi_{\Delta-1}\,d\phi_{\Delta-1}\,d\phi_\Delta
\end{equation}
on this unit sphere. Using this measure, we find that the typical value of the modulus of diagonal elements of $\hat \rho$ is
\begin{equation}
  \sqrt{\langle\,|\rho_{ii}|^2\,\rangle_{d\mu}}=\frac{1}{d_j}, 
\end{equation}
which is the typical size of a coordinate component of a vector in $d_j$-dimensional space. Likewise, the typical value of the modulus of the off-diagonal elements is
\begin{equation}
 \sqrt{\langle\, |\rho_{i\ne j}|^2\,\rangle_{d\mu}}=\frac{1}{D^{N-1}},
\end{equation}
which is the typical size of the inner product between two vectors in $D^{N-1}$-dimensional space. Clearly, this typical size is exponentially smaller than that of the diagonal elements. Thus, the typical density matrix is ``almost'' diagonal \emph{in any basis} and ``almost admits'' an interpretation as a probability distribution on the basis elements. This purely kinematic observation is, when combined with a concrete model for the interactions between subsystems and their environment, the explanation for the experimentally observed decoherence. 

Moreover, we observe that even a small deviation $\epsilon^\alpha$ of the components $v^\alpha$ of a high-dimensional unit vector to another high-dimensional unit vector $u^\alpha=v^\alpha+\epsilon^\alpha$ rapidly decreases the inner product $\langle u|v\rangle\ll 1$. Hence, even if one starts with sizable off-diagonal matrix elements, a generic time evolution will rapidly decrease these off-diagonal elements to exponentially small values. Therefore, we summarise that decoherence is the reason for the rapid ``almost'' diagonalization of the reduced density matrix of a subsystem that interacts with a macroscopic environment, with this ``almost'' diagonalization occurring in any basis. We thus submit that decoherence is not able to \emph{ein}-select a preferred basis.

Let us compare this statement with the paradigmatic model presented in the einselection programme. A system with $d_j=2$ is prepared in a ``classical'' initial state, i.e., a product state with uncorrelated phases, 
\begin{equation}
  |\psi\rangle=\left(\alpha|0\rangle+\beta|1\rangle\right)\otimes_{i=1}^N\left(\alpha_i|0\rangle+\beta_i|1\rangle\right),
\end{equation}
where normalization of the tensor factors requires $|\alpha|^2+|\beta|^2=1$ and $|\alpha_i|^2+|\beta_i|^2=1,\forall i$. Note that this initial state is very atypical: the parameter space for states like this is $3(N+1)$-dimensional, whereas the parameter space for generic states is $2^{N+2}-1$-dimensional. The motivation for this initial state is that it represents a ``classical'' environment, which, as pointed out in \citet{739}, makes the derivation of a classical world from this initial state a circular argument.
 
Moreover, this model also assumes a very special interaction between the subsystem and its environment, of the form 
\begin{equation}
  \hat{H}_{\text{dec}}=\sum_{i=1}^N g_i\,\,\hat\sigma^3\otimes\hat{\sigma}^3_i,
\end{equation}
where $\hat\sigma^3_i=\hat{\mathbb I}_1\otimes\cdots\otimes\hat{\mathbb I}_{i-1}\otimes\hat{\sigma}^3\otimes\hat{\mathbb I}_{i+1}\otimes\cdots\otimes\hat{\mathbb I}_N$, which is, again, highly fine-tuned, as it represents only $N$ parameters out of the $\frac 1 2\,N(N+1)$ possible two-body coupling parameters. Using these highly fine-tuned choices leads to the evolution of the reduced density matrix
\begin{equation}
  \hat \rho=|\alpha|^2|0\rangle\langle 0|+|\beta|^2|1\rangle\langle 1|+z(t)\left(\alpha\beta^*|0\rangle\langle 1|+\alpha^*\beta|1\rangle \langle 0|\right),
\end{equation}
where $z(t)=\prod_{i=1}^N\left(\cos(2\,\frac{g_i\,t}{\hbar})+i(|\alpha_i|^2-|\beta_i|^2)\sin(2\,\frac{g_i\,t}{\hbar})\right)$, which evolves rapidly to the exponentially small typical value. This decay has been used to argue that decoherence selects the $3$-basis. However, as stressed above with the set of initial conditions, the fact that this argument critically depends on highly fine-tuned conditions, this time for the interaction, renders the derivation of classicality circular \citep{739}. 
 
Having pointed out that unitary QM alone is not able to select a preferred basis, we are forced to accept that the very formulation of a quantum theory of the universe requires a preferred basis in which outcomes are realized. In this regard, perhaps the simplest implementation of a preferred basis is found in the de Broglie-Bohm theory, which we shall briefly describe next.

\subsection{The fate of the basis problem in de Broglie-Bohm theory}\label{subsec:basisproblem}

The de Broglie-Bohm theory is closely related to classical mechanics in the following sense: given a configuration space $\mathcal{Q}$, which we assume to be a manifold locally coordinatized by $q^a$, then de Broglie-Bohm theory is a dynamical system that generates a curve $Q^a(\lambda)$, with $Q^a$ the physical configurations and $\lambda$ an arbitrary parameter. However, while in classical mechanics the configuration $Q^a$ is guided by the momenta $p_a$, which satisfy Hamiltonian equations, de Broglie-Bohm mechanics asserts that the configurations are guided by a (suitably smooth) wave function $\psi(q)$ on configuration space, which satisfies a Schr\"odinger equation. 

Hence, the difference between classical mechanics and de Broglie-Bohm mechanics is that the classical momenta $p_a$ are replaced by a guiding wave function $\psi(q)$. However, the instantaneous reality of both theories is a configuration of beables $Q^a$ and the history of the universe is in both cases described by a succession $Q^a(\lambda)$ of objectively definite instantaneous configurations. It is important to point out that the guiding wave function $\psi(q)$ has no other purpose than guide the configuration $Q^a$ of the universe, which is completely analogous to the role of the momenta $p_a$ in classical mechanics.

Let us give an explicit example of a de Broglie-Bohm model. For this, we shall consider a configuration manifold $\mathcal{Q}$ with a Riemannian structure, i.e., a supermetric $g_{ab}(q)$, and a scalar potential $V(q)$ on $\mathcal{Q}$. Moreover, we will work with the polar split of the wave function, $\psi(q)=R(q) e^{i\,S(q)}$, and define the guidance equation for the physically realized configuration $Q^a$ as
\begin{equation}
 \dot Q^a=g^{ab}(Q)\,S_{,b}(Q),
\end{equation} 
where the dot denotes the derivative w.r.t. the evolution parameter $\lambda$. We close the dynamical system by imposing the Schr\"odinger equation for the wave function
\begin{equation}
  \dot \psi(q)=-\frac 1 2 \Delta_g(q)\,\psi(q)+V(q)\,\psi(q),
\end{equation}
where $\Delta_g(q)$ is the Laplace-Beltrami operator of the supermetric $g_{ab}(q)$.

It is clear that de Broglie-Bohm theory does not have an outcome problem, since it predicts a definite history $Q^a(\lambda)$ for the universe (see \citealp{233}, for the argument in the standard version of the theory). Moreover, using the polar split, it is straightforward to see that most of the degrees of the wave function decouple from the dynamics when the gradient of the quantum potential,
\begin{equation}
  U(q)=-\frac{\Delta_g(q)\,R(q)}{2\,R(q)}\,,
\end{equation}
evaluated at the configuration $Q^a$, is negligible for the integration of the system. If we find ourselves in  a situation in which the quantum force $F_a=U_{,a}(Q)$ is negligible, then the only degree of freedom of the wave function that guides the wave function is $S_{,a}(Q)$, which then plays exactly the role of a classical momentum $p_a$. This allows us to identify all circumstances in which the quantum force is negligible as situations that are well described by a classical system with effective Hamiltonian
\begin{equation}
  H_{\text{eff}}=\frac 1 2 g^{ab}(q)p_ap_b+V(q).
\end{equation}
In \citet{509}, it was shown that decoherence together with local plain wave behaviour of the wave function ensures that the de Broglie-Bohm system evolves classically. Intuitively, one understands that macroscopic observables of subsystems, such as the centre of mass of a cluster of particles, experience, for almost every wave function, a ``random'' force contribution to each individual particle, such that the net quantum force for the macroscopic observable averages to zero, whereas potential forces between clusters of particles add up to macroscopic forces.
  
Thus far, we have only seen why de Broglie-Bohm mechanics is analogous to classical mechanics. However, to be compatible with the real world, we need to explain why standard QM emerges as the effective description of the mutual interaction between inhabitants of a de Broglie-Bohm universe. We shall assume that we can perform a tripartite split of this universe into a subsystem, apparatus (or ``observer'') and environment. To simplify the discussion, let us further assume that we can implement this split as a tripartite decomposition of the configuration space into a product of three configuration spaces 
\begin{equation}
  \mathcal{Q}=\mathcal{Q}_{\text{subsys}}\times \mathcal{Q}_{\text{app}}\times \mathcal{Q}_{\text{env}}\,.
\end{equation}

To model an observer, we assume that it is a physical subsystem whose instantaneous state is given by a point in $\mathcal{Q}_{\text{app}}$. The previously mentioned $A$-basis of this observer is chosen as that basis whose elements are uniquely identified by the configuration, i.e., the wave function support of the $|A_n\rangle$ splits $\mathcal{Q}_{\text{app}}$ into disjoint regions $r_n$, such that the subsystem configuration being in one of these regions corresponds to selecting one of the states $|A_n\rangle$ as a distinguishable outcome. This observer is supposed to have a theoretical model to describe the subsystem and its ``measurement-like'' interactions (of the type described in \eqref{equ:MeassurementInteraction}) that allow them to update their description of the state of the subsystem. To be precise, we shall assume:
\begin{enumerate}
 \item The observer uses an empirical density matrix $\hat\rho_{\text{emp}}$ to model their knowledge of the state of the subsystem. In particular, if the observer has no other information than the outcome encoded in the apparatus configuration being shifted from region $r_0$ (the ``ready'' state) to region $r_n$, then the subsystem is assigned the state $\hat\rho_{\text{emp}}=\hat P_n/\textrm{Tr}(\hat P_n)$, where $\hat P_n$ denotes the projector onto the eigenspace of the observed observable which corresponds to the eigenvalue $n$.
 \item The observer has the ability to turn on quasi-instantaneous measurement-like interactions. We will also assume that they set the relevant configuration $Q_k^0$ into its ``ready'' state, to which we will assign the coordinate value $0$, before turning on the quasi-instantaneous measurement-like interaction. Finally, we shall further assume that this interaction is such that if the universe is in a product state $|n\rangle\otimes|\textrm{ready}\rangle\otimes|\psi\rangle$, where $|n\rangle$ is an eigenstate of the subsystem observable $\hat A$, then the interaction will move $Q_k^0$ to the coordinate value $n$.
 \item The internal dynamics of the observer is such that they can store a large number of positions $Q_k^0$ and turn on many interactions, thereby acquiring new ``measurement outcomes'' from the subsystem. The observer is then able to ``update'' their 
previously gathered information through subsequent measurements, by means of a von Neumann or Lindblad equation to evolve the subsystem state, whereby the updates are attained by projecting the evolved density matrix onto the subsequently determined eigenspaces and normalizing the density matrix thereafter.
\end{enumerate}

Let us next analyse whether the information that the observer acquires about the subsystem is in fact well described by the standard quantum formalism. We should like to point out that there exists a large number of wave functions of the subsystem for which the statement above will \emph{not} hold, since, amongst other obstructions, net resulting quantum forces between entangled subsystems are not bounded. We will therefore identify those regimes in which our assumptions are actually realized and argue for their emergence in the model proposed in \S\ref{sec:dBBPSD} (see, in particular, \S\S\ref{subsec:quantumNbody},\ref{subsec:bornstat}). 

The first step consists in arguing that the single trajectory of the universe is {\it{typical}} w.r.t. the density defined by the modulus squared of the wave function. This is essentially argued for by Valentini's work on a ``quantum H-theorem'' (see \citealp{252,253}, for the original papers and \citealp{738}, for a recent review of the programme), whose underlying idea was already suggested in \citet{743}. This can be understood intuitively by observing that (i) the quantum force $F_a(q)=-\left(\frac{\Delta_g(q)\,R(q)}{2R(q)}\right)_{,a}$ tends to move an individual trajectory away from the nodes of the wave function and into the bulges and (ii) the Born density $\rho=|\psi|^2$ is an attractive equilibrium of the dynamics of de Broglie-Bohm trajectories. Thus, given knowledge about a de Broglie-Bohm configuration $Q^a$, which has had some time to evolve, and in the absence of any further detailed knowledge about the wave function, one cannot do better than to assume that $q^a$ is a typical configuration for the Born density distribution $|\psi|^2$.

Secondly, we shall argue that the conditional wave function of a subsystem, 
\begin{equation*}
\psi_c(\vec q_c):=\frac{\psi(\vec Q_r,\vec q_c)}{\sqrt{\int_{\mathcal{Q}_{\mathrm{subsys}}}\,|\psi(\vec Q_r,\vec q_c)|^2}}\,,
\end{equation*} 
with the configuration of the apparatus being the actualized one, $q_r=Q_r$, where the index $r$ refers to the corresponding region associated with the apparatus coupled with the subsystem, becomes an effective wave function. Generically, we can approximate the de Broglie-Bohm wave function with arbitrary precision by sums
\begin{equation}
    \psi(\vec q_r,\vec q_c)=\sum_i \psi^{(1)}_i(\vec q_c)\psi^{(2)}_i(\vec q_r)\,,
\end{equation}
which can be evaluated at $q_r=Q_r$ to give $\psi_c(\vec q_c)=\sum_i a_i\,\psi^{(1)}_i(\vec q_c)$, where the expansion coefficients are $a_i=\psi^{(2)}_i(\vec Q_r)$, such that the time evolution of the conditional wave function becomes
\begin{equation}
    \psi_c(\vec q_c;t)=N(t)\,\sum_i a_i(t)\,\psi^{(1)}_i(\vec q_c;t)\,,
\end{equation}
where $N$ denotes the normalization of the conditional wave function.

Next, we will assume that there is no effective interaction between the subsystem and the rest of the $N$-body system, i.e., the Hamiltonian is very well approximated by $\hat H=\hat H_c\otimes \hat{\mathbb I}_r+\hat{\mathbb I}_c\otimes \hat H_r$. Clearly, at this stage, one cannot identify $\hat H_c$ with the effective Hamiltonian of the subsystem, due to the evolving amplitudes $a_i(t)N(t)$. However, if there are interactions between the subsystem and the environment that ensure that the supports of the environment wave functions $\psi^{(2)}_i(\vec q_r)$ are effectively disjoint in the configuration space, then all $a_i(t)$ vanish approximately, except for a single one, which we will call $a_0(t)$. Intuitively, the physical justification for this claim stems from the fact that we are dealing with a subsystem with a finite number of effectively accessible states and an environment with a large number of degrees of freedom, whereby the environment configuration is free to explore a great many environment dimensions. Hence, after evolution into a regime where the supports of the $\psi^{(2)}_i$ are effectively separated, one is left with 
\begin{equation}\label{equ:DecohSubsyst}
    \psi_c(\vec q_c)\approx \psi^{(1)}_0(\vec q_c)\,.
\end{equation}
This wave function satisfies the effective Schr\"odinger equation
\begin{equation}
    i\partial_t \psi_c = \hat H_c\,\psi_c
\end{equation}
for any non-interacting Hamiltonian $\hat H=\hat H_c\otimes \hat{\mathbb I}_r+\hat{\mathbb I}_c\otimes \hat H_r$\,. 

Based on this framework, one is able to formulate statistical tests to investigate whether the recorded outcomes associated with a typical trajectory are consistent with the predictions of QM. Let us have a closer look at this key issue. Assuming formation of well-defined subsystems within the universe, consisting of a quantum subsystem and a measuring apparatus, the claim of which will be justified more explicitly in the numerical analysis of the solutions of our model in \S\ref{subsec:numerical}, we shall assign the configuration space $\mathcal{Q}_{\text{subsys}}$ to a typical subsystem, and an effective Hilbert space $\mathcal{H}_{\mathcal{Q}_{\mathrm{subsys}}}$. To model our knowledge of the quantum subsystems, we will assume an epistemic (empirical) density matrix $\hat{\rho}_{\text{emp}}$. Next, assuming the effective Hamiltonian $\hat H_c$ for the subsystem, our original question of whether the information that the observer acquires about the subsystem accords well with the empirical content of standard QM translates into the following two questions:

\begin{enumerate}
\item Does our knowledge of the subsystem conform to Born statistics, meaning that for any effective observable $\hat A_c$ the relation
\begin{equation}
\langle \hat A_c \rangle =\textrm{Tr}\left(\hat{\rho}_{\text{emp}}\,\hat{A}_c\right)
\end{equation}
holds?

\item Does the subsystem state obey the unitary dynamics given by von Neumann equation
\begin{equation}
  i\,\dot{\hat{\rho}}_{\text{emp}}=\left[\hat H_{c},\hat\rho_{\text{emp}}\right]?\,,
\end{equation}
\end{enumerate}
where ``dot'' at this stage refers to derivative w.r.t Newtonian time. These two questions must be carefully addressed in any ``fundamental'' quantum model, including our de Broglie-Bohm model (\S\ref{sec:dBBPSD}, where we will describe the unique evolution of the universe and, hence, the evolution of subsystems once they form and effectively isolate). The first question is whether Born rule holds in the model, and the second one is whether probabilities follow the unitary evolution dictated by Schr\"odinger equation.

In the standard de Broglie-Bohm theory, both of these questions are answered in the affirmative. The first one is true because the Born rule is either i) posited as a separate principle for the initial probability distribution, or ii) dynamically derived (see below for a concise account of both stances). The second one also readily follows from the guidance equation and so-called ``equivariance'' of the conservation laws for $|\psi|^2$ distribution and an arbitrary probability distribution $\rho$ in configuration space, as originally noted by Bohm in his original papers. Unlike the standard formulation, in our alternative de Broglie-Bohm model the proof is more demanding, because the unique dynamics of the whole universe is given intrinsically, whereby all probability distributions are defined objectively w.r.t effectively isolated subsystems at a later stage of the evolution (see below).

There are two major camps regarding the understanding and derivation of Born statistics and, hence, the empirical content of QM in the de Broglie-Bohm framework from some fundamental principle(s). One is the ``typicality'' argument proposed in \citet{222}, in which the typicality of Born statistics is justified on grounds of the typicality of the initial state of the whole universe w.r.t the typicality measure $|\psi|^2$ for the whole universe. The competing camp aims to show that Born statistics dynamically ``evolve'' from non-Born statistics for subsystems, a process widely known as ``dynamical relaxation'' to quantum equilibrium from some quantum non-equilibrium initial state. Valentini's ``quantum H-theorem'', already mentioned above, is perhaps the best worked-out example of this approach. Both of these approaches have some advantages and shortcomings. We are more interested in taking a unified stance and combining both of them, but with more emphasis placed on dynamical relaxation.

Given the non-standard character of our de Broglie-Bohm model, the question about the holding of Born statistics is a subtle matter. Let us make it clear what it means for Born rule to apply in our model. We have defined above tests of quantum mechanics as the following series of procedures: i) preparation of an (almost) autonomous subsystem by a first interaction with a first persisting and confined device, ii) subsequent evolution thereof, and finally iii) measurement, through a second interaction with a second persisting and confined device. Given that the outcomes of both preparation and measurement must be recorded in the same persistent medium in order to perform tests, it is thus reasonable to define ``apparatus'' to encompass both the preparation and the measurement devices. We therefore assert that quantum properties are tested accurately when a confined apparatus persists over a long time. This is then the regime in which we need to show that (the empirical content of) standard QM emerges as an effective description of our model. In light of this, we will consider appropriate regimes in which bounded subsystems form and follow a guidance-driven effective de Broglie-Bohm dynamics (\S\ref{subsec:bornstat}). We shall then argue for the emergence of Born statistics for these subsystems.

\section{de Broglie-Bohm Pure Shape Dynamics Model}
\label{sec:dBBPSD}

As pointed out in the introduction, by now it has clearly been established that SD is empirically indistinguishable from GR. Moreover, in \S\ref{subsec:basisproblem} we have presented a review of the known arguments that show that de Broglie-Bohm theory possesses i) a classical limit in which de Broglie-Bohm trajectories become indistinguishable from classical ones and ii) limits in which the information that one semiclassical subsystem acquires about another subsystem is effectively described by the standard quantum formalism. This motivates the construction of a general quantum gravity framework based on a de Broglie-Bohm approach to PSD, by applying the tools developed for the treatment of classical PSD trajectories to trajectories of simple de Broglie-Bohm models on shape space. As emphasised in the introduction, in this paper we shall content ourselves with a much simpler task, to which we turn next.

\subsection{The quantum relational $N$-body system}
\label{subsec:quantumNbody}

The purpose of this section is to construct a quantum model of an $N$-body universe that implements the relational first principles of PSD, which state that there is no absolute notion of scale or duration. In the classical framework, this leads to the postulate that the universe is described by a pure (i.e., unparametrized) curve in shape space, whereby the equations of state of this curve, remarkably enough, reproduce the objective predictions of a class of solutions to the standard (Newtonian or Einsteinian) representations, whenever scale and duration are finite. They typically take the form
\begin{equation}
 \begin{array}{rcl}
  dq^a&=&u^a(\phi)\\
  d\phi_A&=&\Phi(q,\phi,\kappa)\\
  d\kappa&=&K(q,\phi,\kappa)\,,
 \end{array}
\end{equation}
where $q^a$ describes the point in shape space, $\phi_A$ the direction of the curve (coordinates of the unit tangent bundle) and $\kappa$ is related to the extrinsic curvature of the curve in shape space. This equation of state describes how the geometric data of the curve changes as one moves along the curve. Crucially, there is no reference to scale or to any time-parametrization of the curve; the description is solely in terms of the geometry of the curve in shape space. Thus, any notion of scale or duration must be deduced from physical rods and clocks that exist within the pure curve in shape space; once this is done, the standard description of the system is recovered. We shall next perform the analogous purging of scale and duration in quantum systems.

Let us first make a simple, yet key observation: the description of a closed quantum universe does not allow for any external measurement, because all measurements are interactions between physical subsystems within the universe. It is thus necessary to include a law that replaces the measurement axiom of standard QM. Because of its intrinsically relational nature, shape space is a natural arena to achieve this. We will use local coordinates $q^a$ to describe shape space and consider a dimensionless kinematic metric $g_{ab}(q)$ therein. For the sake of definiteness, we shall assume a simple (standard) Hamiltonian of the form
\begin{equation}\label{classicalHamiltonianexpr}
 H=-\frac 1 2 g^{ab}(q)\nabla_a\nabla_b+V(q)
\end{equation}
for the wave functions $\psi(q)=R(q)e^{i\,S(q)}$ in shape space and a standard evolution law for the actual shape $Q^a$,
\begin{equation}\label{guidanceeq}
 \left.\dot Q^a = g^{ab}(q)\nabla_b S(q)\right|_{q^a=Q^a}\,,
\end{equation}
where the dot denotes the derivative w.r.t Newtonian (external) time. Some comments about \eqref{guidanceeq}. First, this system is not a standard de Broglie-Bohm model, because the latter is defined on Newtonian configuration space, rather than shape space, and includes Planck constant $\hbar$, unlike \eqref{guidanceeq}. Second, although manifestly scale-invariant (we only consider wave functions $\psi(q)$ and configurations $Q^a$ in shape space) the system is not yet dimensionless and reparametrization-invariant, due to the explicit appearance of an external time variable. 

To obtain the desired dimensionless and reparametrization-invariant description, we will consider the equation of state that describes the pure succession of instantaneous descriptions of the system, which are pairs $(Q^a,\psi(q))$ of de Broglie-Bohm shapes and wave functions in shape space. We will start with the time-parametrized equations of motion of the non-standard system
\begin{equation}\label{parametrizedBohmianeqs}
 \begin{array}{rcl}
  \dot Q^a&=&g^{ab}(Q)S_{,b}(Q)\\
  \dot R(q)&=&-\left(g^{ab}(q)R_{,a}(q)S_{,b}(q)+ \frac{1}{2}R(q)\Delta\,S(q)\right)\\
  \dot S(q)&=&-\left(\frac{1}{2}g^{ab}(q)S_{,a}(q)S_{,b}(q)+V(q)-\frac{\Delta\,R(q)}{2\,R(q)}\right)\,,
 \end{array}
\end{equation}
where $V_T(q)=V(q)-k\frac{\Delta\,R(q)}{2\,R(q)}$ is the total potential, with $V(q)$ being the classical component (shape potential) and $V_{\mathrm{qu}}\equiv-\frac{\Delta\,R(q)}{2\,R(q)}$ its scale-invariant quantum component. A word is in order. Given \eqref{guidanceeq} lacks Planck's constant, a dimensionless coupling $k$ must be introduced to enable us to have a grasp of the relative strengths of the two components. The physical role of $k$ is to determine the Bohr radius of the ground state of approximately isolated two-body subsystems. Given a particular shape $Q^a$ of the universe and a particular universal wave function, one can observe this Bohr radius as the ratio with the size of the universe. So, given a particular point on the curve in shape space, one can read the value of $k$ off the evolution of $d\phi_A$, if $Q^a$, $\phi_A$, $\kappa$ and $R(q)$ (or at least its value and the value of its Laplacian at $q^a=Q^a$) are known beforehand.

Moreover, we use the kinematic metric $g_{ab}(q)$ on shape space to split $S^1_a :=\left.\nabla_a\,S(q)\right|_{q^a=Q^a}$ into directions $\phi_A$ and an additional degree of freedom $\kappa$:
\begin{equation}\label{ukappadefinition}
 \begin{array}{rcl}
  u^a(\phi)&=&\frac{S^1_a}{\sqrt{g^{ab}(Q)S^1_aS^1_b}}\\
  \kappa&=&g^{ab}(Q)S^1_aS^1_b\,,
 \end{array}
\end{equation}
where $u^a(\phi)$ is a unit tangent vector (w.r.t the kinematic metric $g_{ab}$) at $Q^a$ that is determined by the direction $\phi_A$. With these definitions, we are able to purge all dimensions and the time parametrization by imposing the arc-length parametrization condition
\begin{equation}
\label{arc-length}
 \left(\frac{ds}{dt}\right)^2:=g_{ab}(Q)\dot{Q}^a\dot{Q}^b\,,
\end{equation}
which finally allows us to rewrite the system~\eqref{parametrizedBohmianeqs} as an equation of state in shape space\footnote{As emphasised in \citet{737}, the equation of state describes the relative rates of change of the degrees of freedom of the curve in shape space, $\frac{dq^a/ds}{d\alpha_I^a/ds}=\frac{dq^a}{d\alpha^a_I}$, where $\alpha_I^a$ refers to geometric properties of the curve, hence the absence of the arc-length parameter $s$ in \eqref{eqsofstate}.}:
\begin{equation}\label{eqsofstate}
 \begin{array}{rcl}
  d Q^a &=& u^a(\phi)\\
  d \phi_A &=& \frac{\partial \Phi_A}{\partial Q^a}u^a(\phi)-\frac{\partial \Phi_A}{\partial u^a} \left(\frac 1 2 g^{cd}_{,a}(Q)u_c(\phi)u_d(\phi)+\frac 1 \kappa V_{T,a}(Q)\right)\\
  d \kappa &=& -2u^a(\phi)V_{T,a}(Q)\\
  d R(q) &=& -\frac{1}{\sqrt{\kappa}}\left(g^{ab}(q)R_{,a}(q)S_{,b}(q)+ \frac 1 2 R(q)\Delta S(q)\right)\\
  d S(q) &=& -\frac{1}{\sqrt{\kappa}}\left(\frac{1}{2}g^{ab}(q)S_{,a}(q)S_{,b}(q)+V_T(q)\right).
 \end{array}
\end{equation}

In complete analogy with the classical model, this system is interpreted as generating an unparametrized curve in shape space, which now represents the history of the quantum $N$-body universe. Note that the quantum potential term that appears in $V_T(Q)$ is epoch-dependent and varies as the shape evolves. This is because of the original dynamics~\eqref{parametrizedBohmianeqs}.

Finally, we should like to make a couple of remarks. First, the dynamical system \eqref{eqsofstate} differs from the standard de Broglie-Bohm model in one important aspect: it is not simply the quantized PSD Hamiltonian. Had we followed the standard quantization procedure with~\eqref{classicalHamiltonianexpr}, we would have arrived at a Wheeler-DeWitt-like equation on shape space, as the classical Hamiltonian is a first-class constraint because of the reparametrization invariance. The current model, however, allows the wave function, and hence, the quantum potential to evolve.

Second, there is an important conceptual difference with standard de Broglie-Bohm theory that should be stressed, as the explicit construction given in \S\ref{subsec:basisproblem} already makes it clear: whereas the former considers an ensemble of universes with a distribution given by the quantum equilibrium condition, we assume that the system describes the evolution of a single universe. Despite this, our model does nevertheless exhibit all quantum phenomena, as discussed at length in \S\ref{subsec:basisproblem}. 

\subsection{The meaning and role of scale}\label{subsec:scale}

Our universe is successfully described (at least in the non-relativistic classical domain) by the Newtonian potential, which is inversely dependent on an absolute scale in its expression w.r.t a background frame of reference. This has important implications when we attempt to remove scale and project the dynamics onto shape space. One remarkable feature of our universe, whether considered in the Newtonian or Einsteinian frameworks, is its ``expansion''.

Clearly, this expansion of the universe is not real from the point of view of PSD, for there is simply no background structure w.r.t which expansion can be meaningfully defined. However, a dynamical theory on shape space completely devoid of scale (or an equivalent to that) has already been developed, at least in the case of classical particles \citep{419}, and is shown to be empirically untenable. Thus, instead of either directly coupling scale or removing it altogether, we have to identify its role in the evolution of shape degrees of freedom and change our interpretation of it. As mentioned in the introduction, in \citet{712,706} it was shown that the shape representation of scale makes the classical dynamics attractor-driven, and hence, leads to structure formation and a theory of the arrow of time. Structure formation has a clear representation on shape space, as it directly pertains to pure shapes: shapes with clumpy distribution of point particles are said to be structured. Hence, this is how the intrinsic meaning of scale in SD should be thought of: it brings about ``formation of structures'' by making the dynamics subject to attractors. This should be captured in our quantum PSD formalism, without having to introduce an independent external scale. In order to accomplish this, we will follow the procedure laid out in the classical counterpart \citep{737}.

In classical phase space, scale introduces two variables: its value and its conjugate momentum, known as ``dilatational momentum''\footnote{\label{foot_dilatational}
In the standard Newtonian framework, the dilatational momentum is defined as $D = \sum\limits_{i=1}^N \vec{r_i} \cdot \vec{p_i}$, which was originally proposed in \citet{419}.}. By imposing the zero energy constraint, one of them can be solved for in terms of the other variables in the Hamiltonian, leaving us only one additional quantity to be coupled with the system. This can be achieved by modifying $\kappa$ and its equation in~\eqref{eqsofstate} as follows.

First, in the equation for $d\phi_A$ the total potential $V_T(q)$ is scale-free. We want to consider a more general scale-dependent potential homogeneous in the scale variable $L$ of degree $\alpha$. Hence, we make the transformation

\begin{equation}
V_T(Q) \rightarrow L^\alpha V_T(Q)\,,
\end{equation}
and redefine $\kappa$ as

\begin{equation}
\kappa := \frac{g^{ab}(Q)S^1_aS^1_b}{L^\alpha}\,.
\end{equation}
Now the effect of ``absolute'' change in scale (considered in Newton's absolute space) manifests itself in the equation for the new $\kappa$. Generally, we can include an additional component in the RHS of the equation. Finally, we arrive at the following modified dynamical system with an implicit dynamical role of scale:

\begin{equation}\label{eqsofstatescale}
 \begin{array}{rcl}
  d Q^a &=& u^a(\phi)\\
  d \phi_A &=& \frac{\partial \Phi_A}{\partial Q^a}u^a(\phi)-\frac{\partial \Phi_A}{\partial u^a} \left(\frac 1 2 g^{cd}_{,b}(Q)u_c(\phi)u_d(\phi)+\frac 1 \kappa V_{T,b}(Q)\right)\\
  d \kappa &=& -2u^a(\phi)V_{T,a}(Q) + \tilde{K}(\kappa,\alpha,Q,\phi)\\
  d R(q) &=& -\frac{1}{\sqrt{\kappa}}\left( g^{ab}(q)R_{,a}(q)S_{,b}(q)+ \frac{1}{2} R(q)\Delta S(q)\right)\\
  d S(q) &=& -\frac{1}{\sqrt{\kappa}}\left(\frac{1}{2}g^{ab}(q)S_{,a}(q)S_{,b}(q)+V_T(q)\right),
 \end{array}
\end{equation}
for some function $\tilde{K}$ which we ought to specify, which captures the derivative of~$L$.

The system~\eqref{eqsofstatescale} possesses one more dynamical variable compared to the scale-free system~\eqref{eqsofstate} that completely matches our reasoning above, with only the proviso that one should apply the Hamiltonian constraint on the universe and solve it to find $\tilde{K}$, similarly to the classical case, as will be shown in the next section. The simplest quantum version which reproduces its classical counterpart in its proper limit is (see \citealp{737}, for details):

\begin{equation}
\label{tildeK}
\tilde{K}(\kappa,\alpha,Q,\phi) = \mp\alpha\kappa\sqrt{-\left(1+\frac{2\,V_T(Q)}{\kappa}\right)}\\\,.
\end{equation}

Note that in the system~\eqref{eqsofstatescale} all dynamical variables are dimension-free. This follows from the more general argument that the physics of the whole universe in any framework should be dimensionless, with dimensionful quantities being defined solely for the relations between subsystems. In principle, the effect of all variables can be captured through scale-free quantities, as we have exemplified with $\kappa$.

We should like to emphasise, once again, that although the new modified dynamical system~\eqref{eqsofstatescale} is, strictly speaking, \emph{not} a de Broglie-Bohm model, meaning that it is not the de Broglie-Bohm version of the quantized Hamiltonian~\eqref{classicalHamiltonianexpr}, we nonetheless do stipulate the guidance equations~\eqref{ukappadefinition} for the ``initial'' direction and ``initial'' $\kappa$. The motivation for this condition goes back to the original pilot-wave theory proposed by de Broglie in his contribution to the 1927 Solvay Conference \citep{338}, in which the propagating wave ``acts on'' the configuration through the guidance equation.

A related, and important caveat about the system~\eqref{eqsofstatescale} is that it lacks ``equivariance'' for an arbitrary probability distribution. Equivariance means that any probability distribution $\rho(q)$ considered on configuration space obeys the same conservation law as $|\psi(q)|^2$. This condition is pivotal in de Broglie-Bohm theory, as it allows for the so-called ``equilibrium hypothesis'', $\rho(q) \equiv |\psi(q)|^2$, which makes de Broglie-Bohm dynamics and standard QM empirically indistinguishable. This lack of equivariance can be straightforwardly shown as follows.  

From the fourth equation in~\eqref{eqsofstatescale}, we can readily see that

\begin{equation}
d R^2(q) + \frac{g^{ab}(q)}{\sqrt{\kappa}}\left(R^2(q)S_{,a}(q)\right)_{;b} = 0\,.
\end{equation}
This is the conservation law for $|\psi(q)|^2 = R^2(q)$. However, considering an arbitrary epistemic probability distribution for an ensemble of shapes, denoted by $\rho(q)$, its conservation law would be

\begin{equation}
d\rho(q) + g^{ab}(q) \left( \rho(q) \, u_a(\phi) \right)_{;b} = 0\,.
\end{equation}
The lack of equivariance in this model is due to the fact that $\frac{1}{\sqrt{\kappa}} S_{,a}(q)$ and $u_a(\phi)$  no longer obey the same equations in the system~\eqref{eqsofstatescale}, once we have added the extra term $\tilde K$ for consistency with the classical limit, meaning that the guidance principle does not hold. Crucially, this poses no issues, given that both guidance principle and equivariance are needed to produce the statistical predictions of QM for \emph{subsystems} (see \S\ref{subsec:bornstat} for how Born statistics may emerge within our model). Contrary to this, there is little meaning in the wave function of the whole universe. We have simply a unique global dynamics given by~\eqref{eqsofstatescale} that, in principle, includes large-scale gravitational subsystems described by effective Newtonian mechanics, along with smaller-scale quantum subsystems described by QM. Both standard classical and quantum theories are compacted within this single dynamics for the whole universe.

\subsection{The classical relational $N$-body system}\label{subsec:classicalNbody}

As a consistency check, it is instructive to study the classical limit of our model. Fortunately, the de Broglie-Bohm approach allows a simple investigation of this limit: Whenever the evolution of the quantum potential $V_{\mathrm{qu}}=-\frac{\Delta\,R}{2\,R}(Q)$ along the curve becomes negligible, the evolution of the curve in shape space is effectively described by classical equations of motion determined by the full potential $V_T=V+V_{\mathrm{qu}}$. It is a very delicate enterprise to establish general physical conditions that ensure that the evolution of the quantum potential is in fact negligible. Thus, we shall consider a formal classical limit, in which we will simply assume mathematically that the evolution of $V_{\mathrm{qu}}$ may be discarded.\\

Let us consider the following classical Hamiltonian constraint \citep{737}:
\begin{align}
  \label{Hamiltoniancl}
  H&=\frac{1}{2\,L^2}\left(g^{ab}(q)p_ap_b + D^2\right)+L^{\gamma\,-2}V(q)\approx\,0\,,\nonumber \\
   &\equiv \frac{1}{2}\left(g^{ab}(q)p_ap_b + D^2\right)+L^{\gamma}V(q)\approx\,0\,,
\end{align}
where $D$ is the dilatational momentum introduced in footnote \ref{foot_dilatational}, with $\gamma$ being the homogeneity degree in the scale factor $L$ of the full potential, with $V(q)$ standing for the shape potential. The dynamical system associated with the Hamiltonian \eqref{Hamiltoniancl} reads:
\begin{equation}
 \begin{array}{rcl}
  d q^a &=& u^a(\phi)\,,\\
  d \phi_A &=& \frac{\partial \Phi_A}{\partial q^a}u^a(\phi)-\frac{\partial \Phi_A}{\partial u^a} \left(\frac 1 2 g^{bc}_{,a}(q)u_b(\phi)u_c(\phi)
  +\frac 1 \kappa V_{,a}(q)\right)\,,\\
  d \kappa &=& -2u^a(\phi)V_{,a}(q)\mp\gamma\kappa\sqrt{-\left(1+\frac{2\,V(q)}{\kappa}\right)}\,,\\
 \end{array}
\end{equation}
where $\kappa := p^2 L^{-\gamma}$. Thus, this classical system coincides with the quantum model~\eqref{eqsofstatescale} iff $\gamma\equiv\,\alpha$, meaning that both models must have the same homogeneity degree.

To complete the description of the effective dynamical system, we will also provide so-called \emph{ephemeris equations}, (cf. \citealp{737}, \S~3.6, for a brief presentation of this concept), which become 
\begin{align}
  \frac{d}{ds}\ln\,L&=\pm\sqrt{-\left(1+\frac{2\,V(q)}{\kappa}\right)}\,,\nonumber \\
  \frac{d}{ds}\ln\left[\left(\frac{ds}{dt}\right)^2\right]&=-\frac{2}{\kappa}u^a(\phi)V_{,a}(q)\,.
\end{align}

In a nutshell, these ephemeris equations relate the relational description given in shape space to the standard Newtonian notions of distance $L$ and duration $t$, with $s$ being the arc-length parameter (recall \eqref{arc-length}).

\subsection{Emergence of Born statistics}\label{subsec:bornstat}
As we discussed in \S\ref{subsec:scale}, our model does not follow a guidance principle for the whole universe, which is hardly an issue in the most general sense, but we definitely do need to recover QM for subsystems nonetheless. Our approach is to consider the formation of effectively isolated and dynamically decoupled bounded subsystems. Each one of these subsystems will then follow their own dynamical system that is structurally the same as the global system~\eqref{eqsofstatescale}. We can break this down into three conditions to be satisfied in the appropriate regime:

\begin{enumerate}[label=\Roman*.]
\item The classical component of the total force acting on a particular subsystem $S_I$ is dominant and entirely characterised by degrees of freedom within $S_I$ itself.
\item The quantum force from the rest of the universe has a negligible effect on $S_I$. This is essentially the expected effect of decoherence as described in the derivation of equation (\ref{equ:DecohSubsyst}).
\item We assume either a temporal regime in which $\sfrac{\delta L_I}{L_I}\ll 1$, or, more generally, $\delta L_I \approx 0$, i.e., bounded subsystems have formed that asymptotically tend to a stable state with constant size.
\end{enumerate}

In such a regime, whose existence is strongly supported by the numerical results shown in the next section, we effectively end up having decoupled dynamical systems for subsystems,  similar to~\eqref{eqsofstatescale}, but, crucially, with no additional term for the equation of $\kappa$, because the relative size of these subsystems does not change. This is readily seen as follows: first, from \eqref{tildeK} and \eqref{Hamiltoniancl}, $\tilde{K}=-\gamma\kappa\tfrac{D}{p}$; second, in the classical case, $dL=L\tfrac{D}{p}$. As we are interested in bounded subsystems with constant size, $\tfrac{D}{p}\rightarrow 0\longrightarrow\tilde K\rightarrow 0$. This will ensure that the effective wave function for a given subsystem (recall our construction in \S\ref{subsec:basisproblem}) guides the evolution of this subsystem through standard de Broglie-Bohm dynamics. This feature will in turn lead to equivariance, and hence, consistency of the Born rule for a probabilistic distribution for this subsystem. 

\subsection{Numerical analysis of de Broglie-Bohm trajectories}\label{subsec:numerical}

Having analysed the general framework of a quantum version of PSD, along with a concrete model thereof, it is instructive to get further physical insights into the quantum realm by carrying out a numerical analysis of this model, given the formidable mathematical obstacles to find realistic solutions. Let us first make some comments.
 
With the set of equations~\eqref{eqsofstatescale} one can numerically solve for $R$ and $S$, obtain the quantum potential $V_{\mathrm{qu}}(q) = -\frac{\Delta\,R}{2\,R}$,
and use it to solve the equations for the shape variables. However, this path is somewhat obscure, as various conceptual and technical issues, some of which have already been pointed out above, crystallise into a nuanced situation.

First of all, in order to solve the Schr\"odinger equation we need to specify the initial conditions $R_i$ and $S_i$ on shape space. This poses a conceptual difficulty, because there is no physical principle in the framework of PSD that determines the initial condition for the guiding wave, other than simplicity and lack of redundant initial structures (hence, a homogeneous wave function).\footnote{Initial conditions are always determined by experiments in standard practice. However, what we are concerned about is the \emph{justification} of these conditions from the point of view of our principles, not their empirical role. We believe there must be an underlying reason why we see, empirically, these conditions and not otherwise. In other words, once we move from the physics of subsystems, whose conditions are quite controllable, on to the whole universe, the role of initial conditions also possesses a law-like nature.} Hence, as will be explained below, we shall choose the simplest one, namely an initially homogeneous wave function with no structure at all.

Secondly, as discussed after \eqref{parametrizedBohmianeqs}, we shall include a coupling $k$ next to the quantum potential to control the quantum effect. Fortunately, as will be shown in the results, the ultimate qualitative behaviour of de Broglie-Bohm trajectories seems to be generally independent of this coupling. Moreover, given that the Newtonian potential is scale-dependent and homogeneous of degree $-1$, the quantum potential has also been made homogeneous with the same degree. Therefore, the effective potential reads:

\begin{equation}
    V_T(q,L) = -\frac 1 L \left( \text{Com}(q) + k \frac{\Delta\,R(q)}{2\,R(q)} \right),
\end{equation}
where, for convenience, $k=10^{-1}$ and $\textrm{Com}(q) := -\frac{1}{M^{5/2}}\sqrt{I_{cm}} \, V_N$ is the complexity function originally introduced in \citet{712,706}, defined as the Newtonian potential, $V_N$, made scale invariant through the square root of the centre-of-mass moment of inertia, $I_{cm}$, with $M$ being the total mass of the system. As already emphasised a number of times throughout the paper, complexity plays a most key role in the dynamics in shape space, as will be argued for shortly.  

For the numerical analysis, we shall consider the simple 3-particle model with equal masses. The shape of a 3-particle configuration, i.e., a triangle, can be represented by two of its internal angles, and hence, the corresponding space is a two-dimensional compact surface. A straightforward representation of this shape space is as a sphere, known as \emph{shape sphere}, coordinatized by the azimuthal and polar angles ($\phi$, $\theta$) (see \citealp{712,706}, for the details and definition). Complexity is bounded from below, but has three singularities corresponding to the shape of a triangle with two coincident particles. These singularities are the infinitely deep potential wells of the shape potential and, in this representation, they all lie on the equator $\theta = \pi / 2$.

The kinetic metric on the shape sphere in $(\phi,\theta)$ coordinates is

\begin{equation}
g_{ab} = \left(
\begin{array}{cc}
 \frac{\sin ^2(\theta )}{2} & 0 \\
 0 & \frac{1}{2} \\
\end{array}
\right).
\end{equation}
To solve the differential equations, we have made use of Mathematica for the discretization of the time-dependent Schr\"odinger equation\footnote{\url{https://shorturl.at/BFIO0}}. To implement this function, we have discretized the domain $[0,2\pi] \times [0,\pi]$ of the shape sphere variables ($\phi,\, \theta$) into a grid of $62\times62$ points. This is fine enough for our purpose. This way, the discretized PDE turns into a number of coupled ODEs, whose solution has then been assembled into the final solution.

As mentioned above, the initial wave function has been chosen to be a homogeneous function,

\begin{equation}
    \psi_{ini} = 1\,,
\end{equation}

which readily yields an initially vanishing shape momentum through the guidance equation, and an initially vanishing $\kappa$, too. The initial shape is chosen to be close to the pole with minimum complexity (initial $\theta$ is small). Accordingly, the complete set of initial conditions is
\begin{equation}\label{initialconditions}
\begin{gathered}
    (\phi_{ini},\theta_{ini}) = (2,0.2)\,,\\
    \kappa_{ini} = 0\,.
\end{gathered}
\end{equation}

\begin{figure}[h]
\centering
\begin{subfigure}{0.5\linewidth}
\centering
\includegraphics[width=\textwidth]{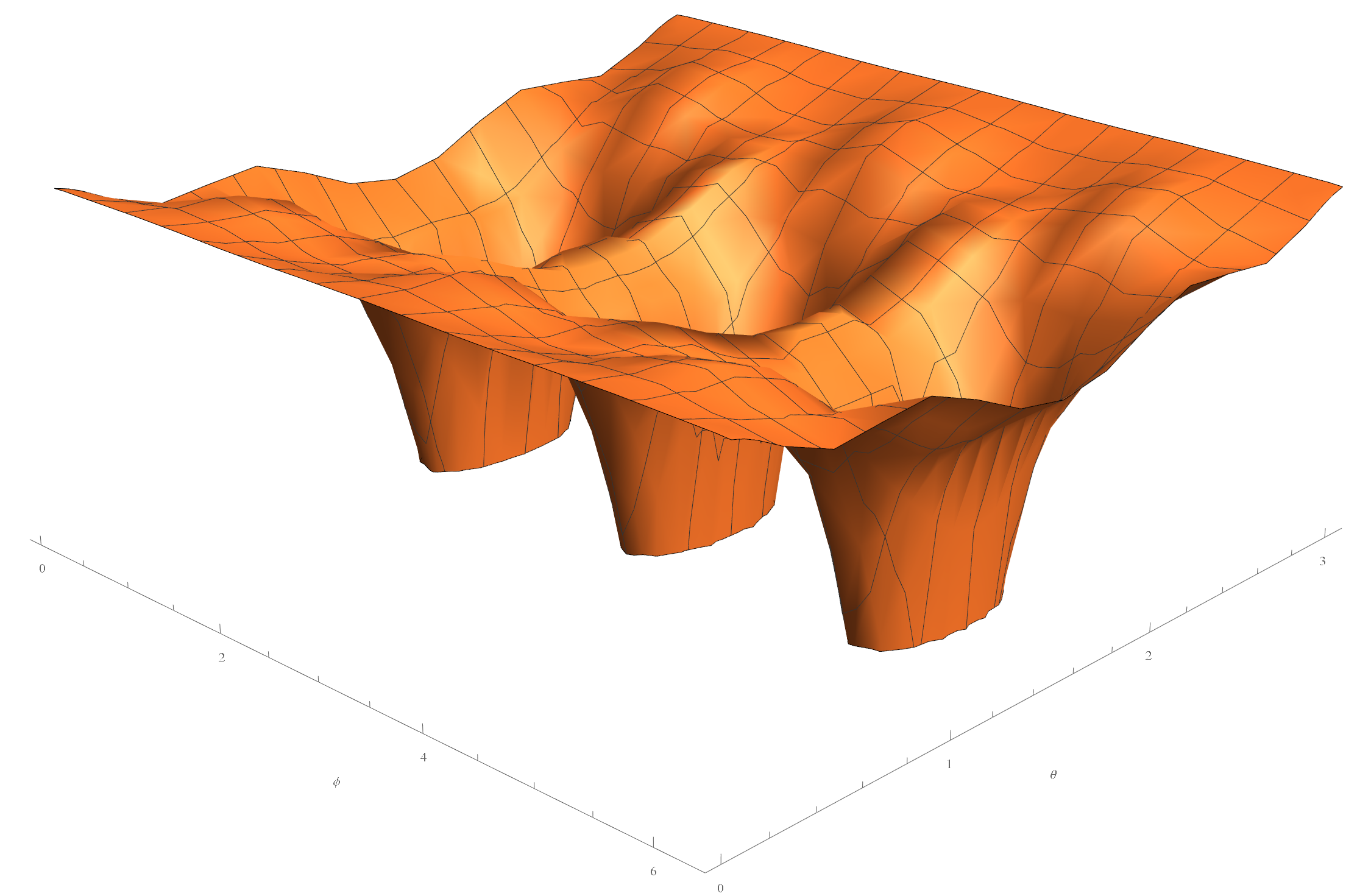}
 \caption{}
 \label{quantumpotentialplot}
\end{subfigure}%
\begin{subfigure}{0.5\linewidth}
\centering
\includegraphics[width=\textwidth]{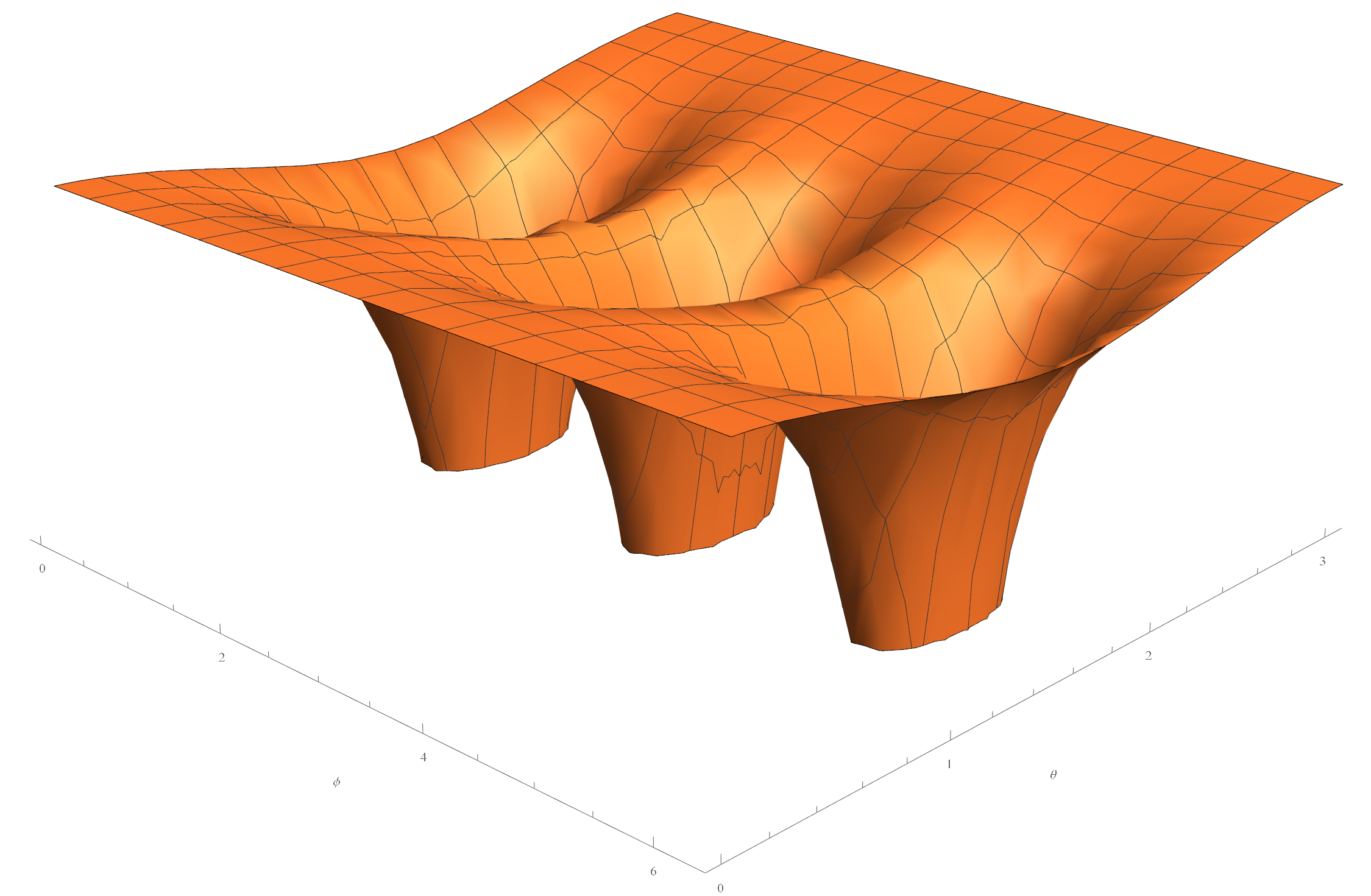}
 \caption{}
\end{subfigure}

\captionsetup{width = 0.9\linewidth}
\caption{\small The left plot (a) is of $V_T(q,1)$ for the 3-body system with equal masses in terms of two parameters representing the shape of the triangle at a specific instant, and the right one (b) is of $-\text{Com}(q)$ (See \citealp{712,706}, for the expression of complexity and definition of parameters). The wells are configurations with coincident points.}
\label{effectivepotentialplot}
\end{figure}

We shall first show the plot of the effective potential cast on shape space and compare it with that of complexity in Fig.\ref{effectivepotentialplot}.

Note that quantum corrections have certainly changed the details of the topography of shape space, but, interestingly, left the singularities of complexity on the equator ($\theta = \pi / 2$) untouched. This, as we will see, ensures the remarkable attractor behaviour even in the quantum model. It should be stressed that this observation is independent of the artificial coupling $k$ inserted by hand: the effect of the purely scale-invariant quantum potential is simply too small to distort the singularities of complexity, although it can potentially change the topography and, hence, the behaviour of trajectories. 

\begin{figure}
  \begin{minipage}[c]{0.5\linewidth}
    \includegraphics[scale=0.4]{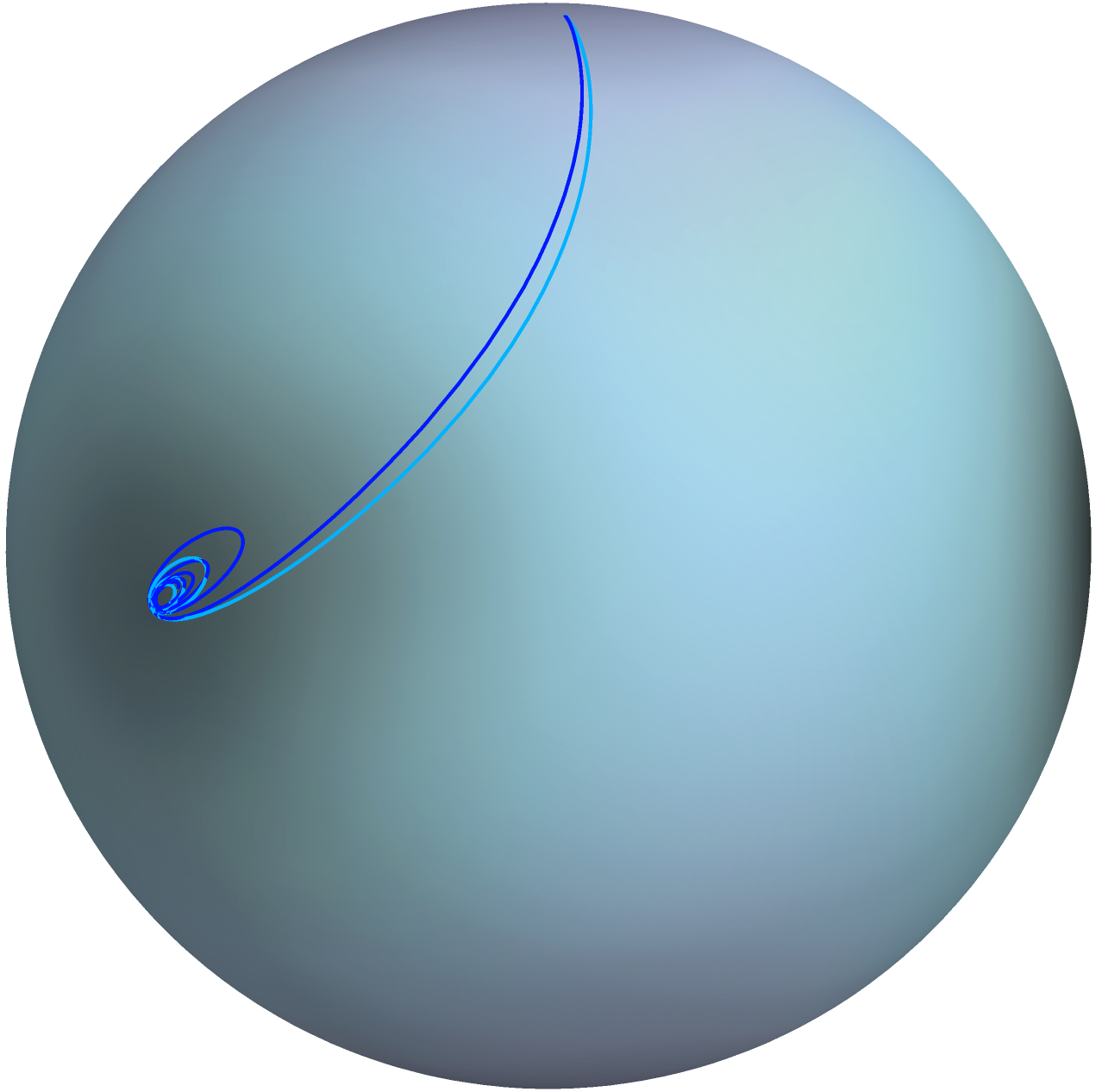}
  \end{minipage}\hfill
  \begin{minipage}[c]{0.4\linewidth}
  \caption{\small The trajectory in light blue is the classical one, whereas the one in dark blue is the de Broglie-Bohm trajectory. Both start with the same initial conditions.}
    
  \label{trajectoriesplot}
  \end{minipage}
\end{figure}

The plot Fig.\ref{quantumpotentialplot} is for an arbitrary instant but not too long after the initial instant. Typically, quantum corrections become suppressed very quickly. With this quantum potential and initial conditions~\eqref{initialconditions}, we can solve the equations of state for shape variables and plot the generated trajectory on the shape sphere (Fig.\ref{trajectoriesplot}). Darker regions on the sphere correspond to higher complexity. Two of the complexity singularities are visible on the front side along the equator. Both trajectories go downwards and get dragged into the front attractor. Remarkably, as already stressed above, they both exhibit the same attractor behaviour in the end, despite the differences in the details of their evolution due to quantum corrections.

Interestingly, we have used many other $\phi$-independent initial conditions for the wave function and, apart from the initial erratic phase and some minute and momentary changes, they all feature the same qualitative results and, hence, the attractor behaviour of the de Broglie-Bohm trajectories. The significance of this result, already emphasised in this paper, lies on the emphasis on complexity as the origin of the arrow of time, as argued for in the classical framework: given the attractor behaviour of the classical solutions for any number of particles, the SD description of classical gravity arguably solves the problem of the arrow of time, as there is no need to posit a special initial condition with low entropy (Past Hypothesis) to justify a macroscopic arrow \citep{706}. Crucially, structure formation, as measured by complexity, already establishes an intrinsic arrow for all solution curves. Thus, our current observation, if true for any generic case, suggests the extension of complexity as the origin of the arrow of time to the quantum realm.

Finally, given the numerical analysis, we find it reasonable to frame the following conjecture: \emph{the Schr\"odinger equation itself considered on shape space possesses attractors}. If true, this conjecture also implies the attractor behaviour of de Broglie-Bohm trajectories in a more general manner. The investigation of this conjecture as well as probing the possibilities it opens are beyond the scope of this paper.

\section{Conclusions}\label{sec:conclusions}

If one aspires to understand the quantum dynamics of the whole universe in a simple realistic framework, free from the notorious conceptual difficulties of standard QM, de Broglie-Bohm theory is a natural candidate, though we should like to emphasise that our commitment to it is rather pragmatic at this stage and may likely be superseded by more complete, perhaps radical approaches currently in progress. Thus, we believe the model put forward in this paper may be a launching pad for a more comprehensive study of ``quantum shape dynamics''. Let us recapitulate the key results derived in this paper.

Standard QM suffers from two major drawbacks, which have seriously impeded progress of research in quantum gravity, namely, its background dependence on absolute spatial and temporal structures. In our model, the first obstacle is easily removed by formulating the dynamical system on shape space rather than Newtonian configuration space. Likewise, the second aspect, which is essentially the source of the so-called ``problem of time'', is overcome by implementing (a relational version of) the pilot-wave framework, in which a unique evolution of the universe is given as a pure, unparametrized, geometrical curve in shape space.

Moreover, this model, as shown in \S\ref{sec:OutcomeProblem}, automatically solves the preferred basis problem by breaking the conceptually problematic unitary equivalence of basis selection of QM and favouring the fundamental space of shapes as our objectively real ontology. 

Finally, given the numerical analysis of the solutions, it is established, at least for the three-body case, that the model gives the expected classical behaviour of the universe asymptotically: quantum dynamics affects the evolution of the universe in the initial phase, but leaves the attractor behaviour of dynamics, and thus, structure formation, intact.

There remain several open problems to consider. A conceptual issue is to look for a physical principle underlying the origin of initial conditions, unlike the case of the current model, where an initially homogeneous potential has been chosen solely on grounds of simplicity and reason. As explained in \S\ref{subsec:numerical}, following Julian Barbour's suggestion, the point with minimum complexity could be an acceptable candidate for the origin of the universe, and time. Thus, we find it absolutely pivotal to keep looking for an all-encompassing theory that provides a sufficiently reasonable explanation for the initial conditions. The questions of ``why these laws and initial conditions'' are becoming increasingly crucial in physics. 

A second major issue regards the nature of the wave function, which might be thought of as an awkward inhabitant of shape space. Admittedly, in our model the wave function is on a par with shapes, which definitely raises doubts about the relational tenets of the whole approach. The reason why it is a very challenging task to get rid of the wave function, through some kind of reduction to shapes, is simple: the propagation of \emph{infinitely} many relevant degrees of freedom of the wave function, which cannot be accommodated by the \emph{finite} number of shapes degrees of freedom. Clearly, then, systems that may be described by a wave function with a finite number of degrees of freedom (modelled by Gaussians, say) will in principle be amenable to the kind of reduction above (however challenging the task may be mathematically). Now then, whether realistic systems, associated with a Gaussian-like wave function, actually exist is ongoing research. A bolder proposal would be to go beyond the simple framework of the model presented in this paper, either by a generalisation or through the implementation of new principles. This, too, is the matter of current research.  
  
Another important issue is the quantum coupling, which we have added to the dynamical system of our model by hand (recall the absence of Planck's constant in our model). Although in \S\ref{subsec:quantumNbody} we provide a representation of this coupling by means of Bohr radii, we currently lack an explanation for its necessary existence in the first place, at least in our model, in order to make a viable theory of the universe. A theoretical explanation for the emergence of this coupling (along with the ensuing appearance of Planck's constant) and, more generally, the emergence of dimensionful base quantities and constants of Nature is a topic for future research.

All in all, the current model does capture the quantum behaviour of the $N$-body system in a relational and background-independent fashion, following PSD tenets. This opens the possibility for exploring a simple, yet powerful alternative framework for quantum gravity. What we present in this paper is perhaps the first step in this direction.

%\newpage

\section*{Acknowledgements}\pdfbookmark[1]{Acknowledgements}{acknowledgements}

Tim Koslowski would like to thank many collaborators for their support of the Shape Dynamics programme. His first and foremost thanks go to Julian Barbour, who has developed many of the conceptual ideas exploited in this research field. Many thanks go also to Henrique Gomes, Flavio Mercati, David Sloan and Sean Gryb for many stimulating discussions about Pure Shape Dynamics. Pedro Naranjo would also like to thank Julian Barbour for his hospitality at College Farm, where P. N.'s interests in relational physics started to develop, as well as for recent discussions. Also, discussions with Sean Gryb and Flavio Mercati are much appreciated. Pooya Farokhi is deeply grateful to Julian Barbour for his support, encouragement, and many valuable discussions. Special thanks go also to Sean Echols and his group at California Polytechnic State University for their help with the development of the numerical program. P. F. is grateful for the scholarship from Bonn-Cologne Graduate School which has facilitated the progress of this research. 

\bibliography{biblio}

\end{document}